\documentclass[conference]{IEEEtran}
\IEEEoverridecommandlockouts
\usepackage{cite}
\usepackage{amsmath,amssymb,amsfonts}
\usepackage{algorithmic}
\usepackage{graphicx}
\usepackage{makecell}
\usepackage{svg}
\usepackage{booktabs}
\usepackage{textcomp}
\usepackage{xcolor}

\graphicspath{ {.} }

\def\BibTeX{{\rm B\kern-.05em{\sc i\kern-.025em b}\kern-.08em
    T\kern-.1667em\lower.7ex\hbox{E}\kern-.125emX}}
\begin{document}

\title{SecDT: A Profile-Based Security Layer for TRDP Communications\\
}

\author{\IEEEauthorblockN{Erlantz Alonso}
\IEEEauthorblockA{\textit{Dept. of Communications Engineering}\\
\textit{University of the Basque Country (EHU)}\\
Bilbao, Spain \\
ealonso074@ikasle.ehu.eus}
\and
\IEEEauthorblockN{Igor Lopez}
\IEEEauthorblockA{\textit{Technology Division}\\
\textit{CAF S.A.}\\
Beasain, Spain \\
igor.lopez@caf.net}
\and
\IEEEauthorblockN{Jasone Astorga}
\IEEEauthorblockA{\textit{Dept. of Communications Engineering}\\
\textit{University of the Basque Country (EHU)}\\
Bilbao, Spain \\
jasone.astorga@ehu.eus}
}

\maketitle

\begin{abstract}
The Train Real-time Data Protocol (TRDP) is widely used on rolling stock but it provides limited native support for cryptographic protection. Furthermore, the multicast traffic profile used in TRDP Process Data to exchange critical information between onboard subsystems makes the introduction of cryptographic protection a challenge. This paper presents a lightweight security layer for secure TRDP communication that implements a number of security profiles built around modern cryptographic algorithms. This additional layer relies on an On-board Key Management System (OKMS) for both security profile negotiation, dynamic key distribution and key lifecycle management. The security profiles allow for cryptographic agility and flexibility, ranging from simple authentication to Authenticated Encryption with Associated Data (AEAD) algorithms. The security profile negotiation procedure guarantees all TRDP End Devices (ED) on a common Communication ID (ComID) share the same security profile and can therefore process each other's messages. A prototype implementation based on mbedTLS and Arm Platform Security Architecture (PSA) was developed and evaluated. Experimental results demonstrate manageable overhead, suitable for the real-time and time-sensitive communication found on rolling stock.
\end{abstract}


\section{Introduction}
Increasingly connected and automated mobility (CAM) systems heavily depend on communication networks to support operational, safety-critical and maintenance functions. Within the railway domain, these requirements are addressed by the Train Communication Network (TCN), standardized in IEC 61375. The Train Real-time Data Protocol (TRDP) is a key component of this architecture, enabling deterministic data exchange between on-board devices and systems and supporting the evolution towards more connected and automated railway environments.

Traditional rolling stock communications relied heavily on physical isolation and secure physical environments to remain secure themselves. However, modern deployments which interact with external systems and remote digital infrastructure introduce new attack vectors that cannot be addressed by legacy security measures. Moreover, the railway sector is facing a demand for increased target security levels, following standards such as IEC 62443, TS50701 and IEC 63452, underlining the necessity for secure and scalable communication mechanisms.

Existing security mechanisms, such as Transport Layer Security (TLS) and network-level protection methods, can provide communication security. These, however, are not always well suited to real-time IP multicast communication as different applications may require different security guarantees or have different performance capabilities. Some communications may only require message authentication and integrity protection, whereas other communications require confidentiality as well. A flexible approach capable of adapting security guarantees to the requirements of individual communication flows and capabilities of the devices in those flows is therefore desirable.

\begin{figure}[ht]
    \centering
    \includegraphics[width=0.65\linewidth]{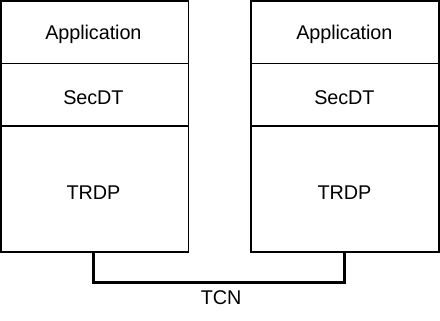}
    \caption{TCN with SecDT protected TRDP protocol stack.}
    \label{fig:Stacks}
\end{figure}

To tackle these challenges, this paper presents Secure Data Transmission (SecDT) for TRDP, a security profile-based security layer for TRDP communications, which rests above TRDP as indicated in Figure \ref{fig:Stacks}. By allowing the security mechanisms to be defined per-communication, SecDT maintains the balance between security requirements and computational overhead. SecDT integrates with a centralized On-board Key Management System (OKMS) responsible for distributing security information to participating devices.

The SecDT security layer does not modify regular TRDP communication or its underlying networking control, meaning a SecDT capable device is backward-compatible with communications that do not make use of SecDT. TRDP End Devices (ED) not capable of performing the proposed security functions can still send and receive application data in insecure ways on Communication IDs (ComID) otherwise populated by SecDT capable devices which are communicating using the SecDT layer.

In order to remain future-proof, communication with the OKMS is achieved through a REST API protected by mutual TLS (mTLS). As Train Control Networks (TCN) become the targets of more advanced threats, future TLS developments to face said threats can quickly be implemented, with no further modification to SecDT core functions.

The main contributions of this work are the design of a profile-based security layer for TRDP, centralized security profile and key management, and a performance evaluation on embedded hardware.

This paper is structured as follows: Section \ref{related_work} presents the related work, Section \ref{proposed_secdt_design} describes the context in which SecDT is expected to operate and how it provides security within it, Section \ref{key_management} covers the OKMS and how TRDP EDs communicate with it, Section \ref{performance_evaluation} presents relevant performance details, metrics and benchmark results and Section \ref{conclusions_and_future_work} summarizes the objective, design and achieved results, followed by a brief exploration of possible evolution paths for this work.

\section{Related Work}
\label{related_work}
Attempts to provide TRDP, and conceptually similar industrial protocols, with modern security measures have been carried out following differing methodologies.

MQTT, a lightweight protocol widely used in Internet of Things (IoT) and industrial applications, much like TRDP follows a publisher-subscriber approach to group communication. It employs brokers instead of IP multicast communication, and it relies on TLS to secure communications between clients and brokers \cite{alharbi_application_2025}. Relying on a broker, however, makes MQTT vulnerable at a single point of failure and introduces latency. TRDP relies on IP multicast communication to meet real-time requirements, making TLS-based channel protection unsuitable and highlighting the need for message-oriented security.

Some previous work has focused on detecting attacks against TRDP instead of preventing unauthorized messages from being processed as legitimate \cite{mihhail_sokolov_towards_2024}. Intrusion Detection Systems (IDS) provide a valuable defensive layer, but operate only on traffic that has already been transmitted and may be processed as legitimate by the receiving device. IDS solutions may be extended into Intrusion Prevention System (IPS) based solutions, which can trigger mitigations. Such approaches remain fundamentally reactive, as malicious traffic must be detected before any action can be taken. SecDT instead focuses on proactive protection, through cryptographic mechanisms, to prevent unauthorized traffic from being accepted in the first place.

Efforts have been made to protect TRDP traffic through cryptographic mechanisms, protecting integrity and performing authentication for real-time TRDP communication and also protecting data confidentiality for certain TRDP information exchanges \cite{_security_2025}. While sharing objectives with SecDT, \cite{_security_2025} focuses on a single authentication mechanism for multicast communication. SecDT aims to provide configurable security mechanisms through cryptographic agility instead.

OPC Unified Architecture (OPC UA) in its PubSub security and SecDT implement similar mechanisms, allowing different levels of protection according to application requirements, and multicast support through message-oriented protection \cite{cryptoeprint:2025/148}, \cite{opc_foundation_opc_2025}. While OPC UA PubSub demonstrates the possibility of such a security mechanism in industrial systems, SecDT differs by targeting existing TRDP systems, their compatibility and integration requirements.

To summarize, while some security measures have focused on protecting TRDP, a lack of flexibility may limit adaptation to future requirements. Security mechanisms exist in other industrial communication systems that share core functionality with SecDT, but they are not applicable to existing TRDP deployments without significant architectural modifications. SecDT addresses these security concerns while maintaining existing TRDP communication behavior.

\section{Proposed SecDT Design}
\label{proposed_secdt_design}

\subsection{TRDP Messages}
TRDP itself focuses on deterministic real-time data exchanges. A TRDP Process Data (PD) telegram is used for periodic exchanges, often using IP multicast communication, identified by a ComID. A PD telegram consists of a fixed header, containing basic communication and source identification information, and the payload, made up of arbitrary application data. This work focuses on TRDP PD telegrams: their strict timing requirements and use of IP multicast cause the most complexity amongst all TRDP communication types.

Since no security mechanisms are supported by TRDP, SecDT addresses this limitation by encapsulating the application data before inserting it into the TRDP payload field.

\subsection{Threat Model}
SecDT assumes an attacker capable of observing, recording and injecting traffic on a TCN. An attacker may attempt message modification, spoofing and replay attacks, albeit without knowledge of the cryptographic keys used to protect communication.

The OKMS, TRDP ED certificates and used cryptographic algorithm implementations are assumed to be trustworthy and uncompromised. The TLS secure channel employed for key distribution is assumed to provide authentication, integrity and confidentiality protections and to remain uncompromised.

DoS attacks against the OKMS or network infrastructure are outside the scope of this work.

\subsection{SecDT Message Structure}
\begin{figure}[ht]
    \centering
    \includegraphics[width=0.5\linewidth]{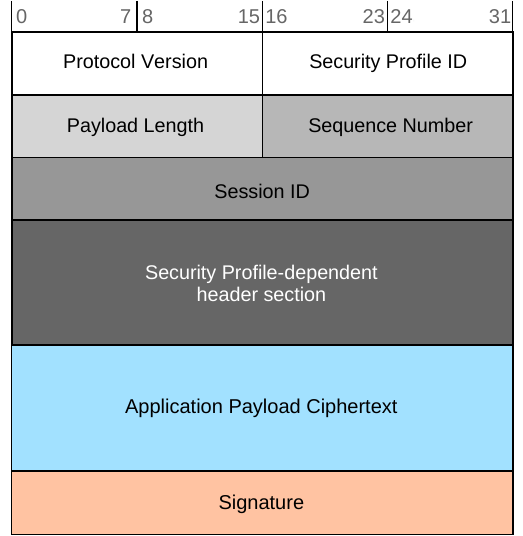}
    \caption{Diagram showing the SecDT message structure and header. The fields not filled in white represent the fields involved in authentication.}
    \vspace{-5pt}
    \label{fig:HEADER_Diagram}
\end{figure}

SecDT uses a two-part header, as seen in Figure \ref{fig:HEADER_Diagram}, to communicate what security measure types are used by any given message. The first part of the header includes basic information, such as the protocol version, used security profile identifier, length of the application payload, sequence number and Session ID. The second part, labeled "Security Profile-dependent header section" in Figure \ref{fig:HEADER_Diagram}, varies in content depending on the used security profile but not in length, as it is always 48 Bytes long. Usually, this part transports the SHA-256 hash of the key or keys used and the Initialization Vector (IV) used, if any.

\subsection{Security Profiles}
Different communication types exhibit different requirements. Diagnostic messages may require only integrity protection, while maintenance traffic and proprietary system data may additionally require confidentiality protection. Furthermore, certain devices may not be able to take on a large computational overhead as required by encryption mechanisms. Being able to adapt to any specific TCN requirements and capabilities has been a key factor for designing a system based on security profiles.

The security profiles in Table \ref{tab:SecProf_list} have been defined in SecDT in order to address the security challenges TRDP communications face.

Amongst these, HMAC-SHA-256 and HMAC-SHA3-256 profiles provide authentication and integrity without confidentiality. GMAC also provides integrity protection in a way that may benefit from AES hardware acceleration when such functionality is available.

AES-CBC+HMAC and AES-GCM profiles additionally provide confidentiality: AES-GCM provides AEAD while AES-CBC+HMAC separates encryption and authentication operations.

\begin{table}[ht]
    \centering
    \begin{tabular}{cccc}
        \thead{\textbf{Security}\\ \textbf{Profile ID}} & \thead{\textbf{Name}\\ \textbf{or Description}} & \thead{\textbf{Auth.} \\ \textbf{and Integrity}} & \thead{\textbf{Confiden-}\\ \textbf{tiality}} \\
        \midrule
        0x0000 & HMAC-SHA-256 & Yes & No \\
        0x0001 & \makecell{AES-256-CBC \\ + HMAC-SHA-256} & Yes &  Yes \\
        0x0002 & GMAC-256 & Yes  & No \\
        0x0003 & AES-256-GCM & Yes & Yes\\
        0x0004 & HMAC-SHA3-256 & Yes & No\\
        0x0005 & \makecell{AES-256-CBC \\ + HMAC-SHA3-256} & Yes & Yes \\
        0xFFFF & None & No & No \\
    \end{tabular}
    \caption{Security profiles defined in SecDT.}
    \vspace{-15pt}
    \label{tab:SecProf_list}
\end{table}

\subsection{Associated Data Authentication}
\begin{figure}[ht]
    \centering
    \includegraphics[width=0.5\linewidth]{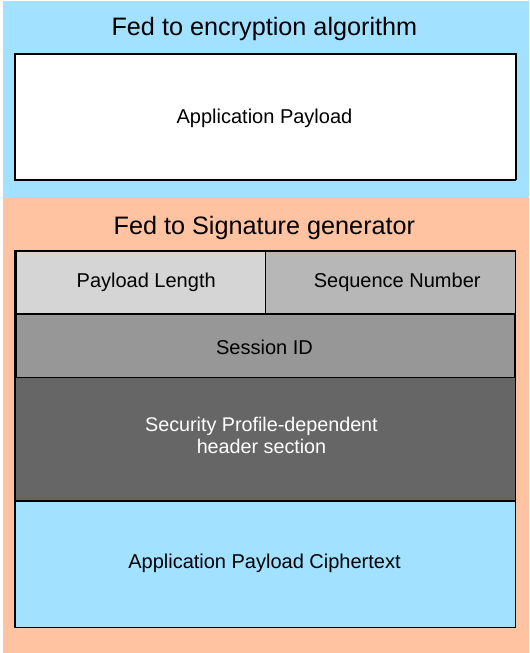}
    \caption{Authentication data inputs (gray, blue) structure and output (orange). Security profiles that do not encrypt replace the blue sections with the "Application Payload" field directly.}
    \vspace{-5pt}
    \label{fig:AEAD_Diagram}
\end{figure}

In order to also guarantee replay protection, SecDT makes use of associated data authentication mechanisms. The application payload, encrypted or not, is prepended by the payload length, sequence number, Session ID and the Security Profile-dependent header section before signing. Figure \ref{fig:AEAD_Diagram} illustrates how this mechanism is used in a security profile that also guarantees confidentiality.

Protecting the OKMS-assigned Session ID and sequence number, alongside receiver-side rejection of previously accepted sequence numbers within the same Session ID, provides replay protection. Protecting the payload length prevents manipulation of message size metadata, ensuring receivers process the expected amount of application data.

\subsection{Backward Compatibility}
Due to the difficulty in achieving SecDT adoption by all rolling stock subsystem suppliers, it has been designed to be backward-compatible. This allows for gradual adoption, depending on the cybersecurity impact of each subsystem.

A TRDP ED incapable of implementing SecDT may still send and receive messages on a ComID protected by SecDT, but it will do so with no security measures of any kind. Legacy compatibility is preserved without modification to the underlying TRDP protocol, it does, however, impose some limits on the use of SecDT itself: the security profiles in a ComID where backward-compatibility is necessary are limited to the security profiles only capable of protecting message integrity and authentication.

Backward compatibility requires legacy applications to ignore the security information and extract the application payload from the fixed offset and trailing signature structure used by SecDT.

\section{Key Management}
\label{key_management}
The TCN is extended by introducing an OKMS. The OKMS is aware of all TRDP EDs and what ComIDs they are allowed to use to communicate. In order to determine what security profile and key or keys each ComID uses, a short initial negotiation is used. Once the security profile and key or keys have been set, the OKMS may update, modify or revoke keys from TRDP EDs via the HTTP2 server-push Server Sent Events (SSE) mechanism.

\begin{figure}[ht]
    \centering
    \includegraphics[width=0.8\linewidth]{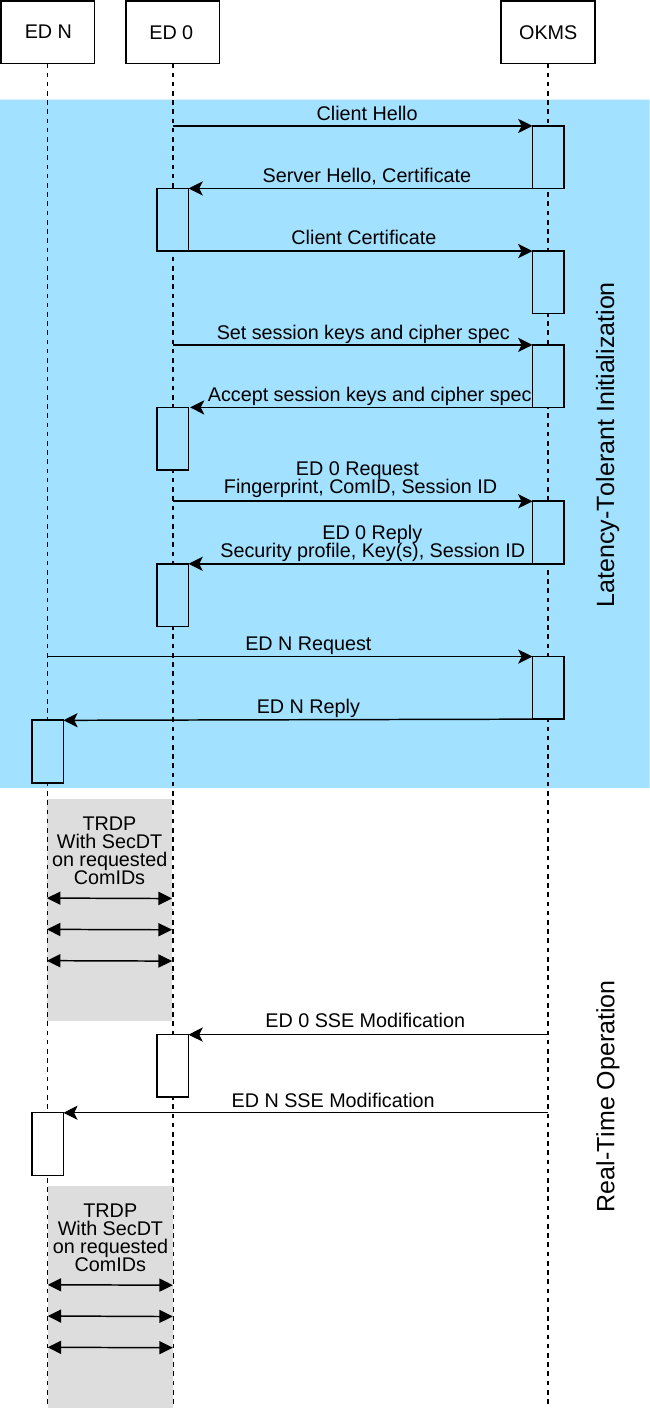}
    \caption{SecDT communication sequence diagram.}
    \vspace{-10pt}
    \label{fig:Sequence_Diagram}
\end{figure}

Figure \ref{fig:Sequence_Diagram} contains a sequence diagram illustrating how systems implementing SecDT interact with an OKMS.

\subsection{OKMS mechanisms}
TRDP EDs may communicate with an OKMS through regular request-reply exchanges or by requesting an SSE flow: the OKMS REST API contains a different endpoint for each communication method.

\begin{figure}[ht]
    \centering
    \includegraphics[width=0.6\linewidth]{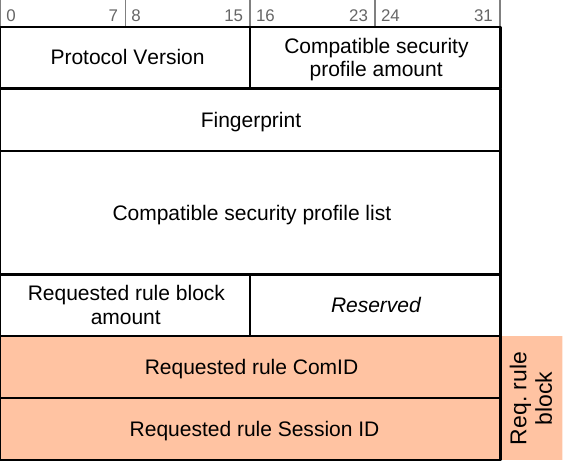}
    \caption{OKMS Request message structure.}
    \vspace{-5pt}
    \label{fig:OKMS-Request}
\end{figure}

EDs send a POST request to either endpoint following the structure in Figure \ref{fig:OKMS-Request}: A header section identifies the TRDP ED and its capabilities, and a number of requested rule blocks communicate which ComIDs the requesting ED needs to securely communicate on. If a request is an update request, the previously generated Session ID is communicated too, if not, the Session ID field shall be set to 0. The ED then listens for either a single reply or a flow of SSE, depending on the endpoint used.

\begin{figure}[ht]
    \centering
    \includegraphics[width=0.6\linewidth]{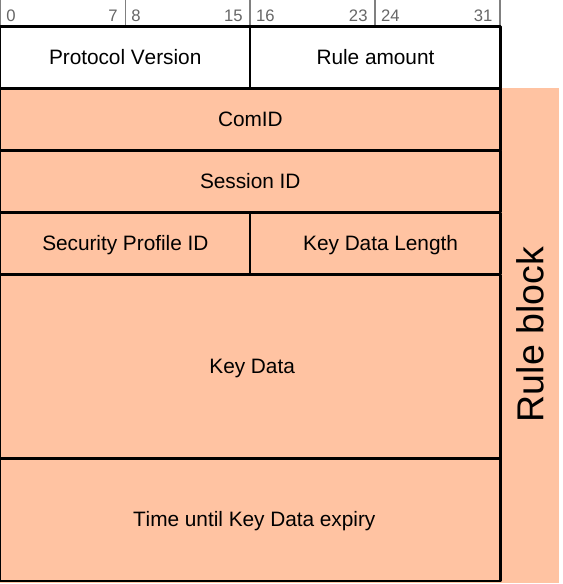}
    \caption{OKMS Reply and SSE message structure.}
    \vspace{-5pt}
    \label{fig:OKMS-Reply}
\end{figure}

The OKMS generates a reply or flow of SSE following the structure in Figure \ref{fig:OKMS-Reply}. The OKMS may be configured to reply with a series of rule blocks, no longer than the list of requested rule blocks, or with an individual reply for each generated rule block. In each rule block, Session IDs are generated if needed, security profiles are assigned and keys and expiry times are communicated.

\subsection{Security Profile Negotiation}
OKMS negotiation must remain secure in order to guarantee that the keys that will later be used by SecDT are not compromised. For this reason, the OKMS exposes a HTTP/2 REST API, only accessible through an mTLS secure channel where both the TRDP ED and OKMS have validated each other's certificates. The method of certificate generation and assignment lies beyond the scope of this security layer, any systems integrator wishing to use SecDT may use the system they deem appropriate.

Once a secure channel has been established, a TRDP ED makes a request using the structure previously described. Upon receiving a request, the OKMS validates the identity, authorization and compatibility of the requesting device before replying with the correct structure, fetching or generating key data and key expiry information.

The OKMS is solely responsible for selecting the security profile assigned to each ComID. During negotiation TRDP EDs advertise the security profiles they support. To prevent security profile downgrade attacks, this information is used only to determine device compatibility and cannot influence the OKMS or the security profile configured for each ComID.

After successful negotiation, a TRDP ED will have received a list of which ComIDs it can communicate securely on, what security profiles to use on each of those along with what keys to use, what Session ID to use on each ComID and when this information will expire.

\subsection{Security Profile Modification}
The OKMS can modify, update or revoke security information at any time through SSE. SSE is an HTTP/2 mechanism to allow for push communication. For this purpose, the OKMS sends a message following the same structure as the OKMS reply.

Upon receiving an SSE, a TRDP ED must implement the changes in a make-before-break approach. During this interval receivers accept messages protected with both the updated and previous security materials while transmitters only use the updated material. Receivers that have not yet been updated but receive data using unknown security materials poll the OKMS for updates.
Revoking security material may instead follow a break-before-make approach as more security material is not expected.

\subsection{Key Lifecycle}
A key or set of keys, whichever fits the security profile, is generated by the OKMS at the end of the negotiation process. Additionally, a new Session ID is generated whenever the security configuration associated with a ComID changes, such as initial distribution, key renewal or security profile modification.

Keys are distributed alongside an expiration window, indicated in milliseconds, before the end of which the key or keys must be renewed. A window is used in place of a fixed timestamp because not all TRDP EDs are predictably equipped with real-time clocks (RTCs) nor can it be expected that all RTCs on a TCN will be adequately synchronized with one-another.

A TRDP ED shall renew security information before expiration, although the OKMS may also perform updates through SSE. If the OKMS does not reply to a key renewal request and is therefore left to expire, the TRDP ED will cease to send secure data using the expired security configuration, but may still receive messages that fit the now expired security configuration.

\section{Performance Evaluation}
\label{performance_evaluation}
SecDT is designed to provide security on embedded systems, and as such it is expected to have a minimal performance footprint. Raspberry Pi 4 computers acting as transmitting and receiving TRDP EDs were used as performance testbenches.

\subsection{Implementation}
A prototype of SecDT has been developed to evaluate the feasibility of profile-based security for TRDP. This implementation builds upon TCNOpen TRDP Light 3.0.0.0. The cryptographic functions are provided by the mbedTLS 3.6.6 library and Arm Platform Security Architecture (PSA). Communicating with the OKMS is handled by curl.

Evaluation of large-scale key distribution behaviour is not the objective of this work, however, an OKMS implementation serves as a functional prototype handling security profile negotiation, key distribution and key lifecycle management as required to evaluate SecDT.

\subsection{Overhead}
SecDT employs a statically-sized 60 Byte header, paired with the already large 48 Byte TRDP PD header, overhead can be a significant factor for communications with large amounts of small messages.

The decision to keep a statically-sized large header was made to ease the backward-compatibility mechanism mentioned in this document, the benefits of which outweigh the performance loss caused by substantial overhead.

\subsection{CPU Time Requirements}
Evaluation is necessary to determine whether SecDT and the security profiles defined remain compatible with the timing requirements imposed by TRDP communication, and to specify the processing-cost difference between said security profiles.

The evaluation compares all security profiles defined in Table \ref{tab:SecProf_list} against an insecure baseline implementation performing only payload-copy operations.

CPU processing time was measured during transmission and reception on Raspberry Pi 4 devices, making use of no hardware-based cryptographic acceleration. The benchmark measures only SecDT processing time: network handling and operating-system scheduling were excluded to isolate the cost of the security layer itself. OKMS negotiation was also omitted as it is performed only once during the latency-tolerant initialization process.

\begin{figure} [ht]
    \centering
    \includegraphics[width=0.85\linewidth,height=0.95\textheight,keepaspectratio]{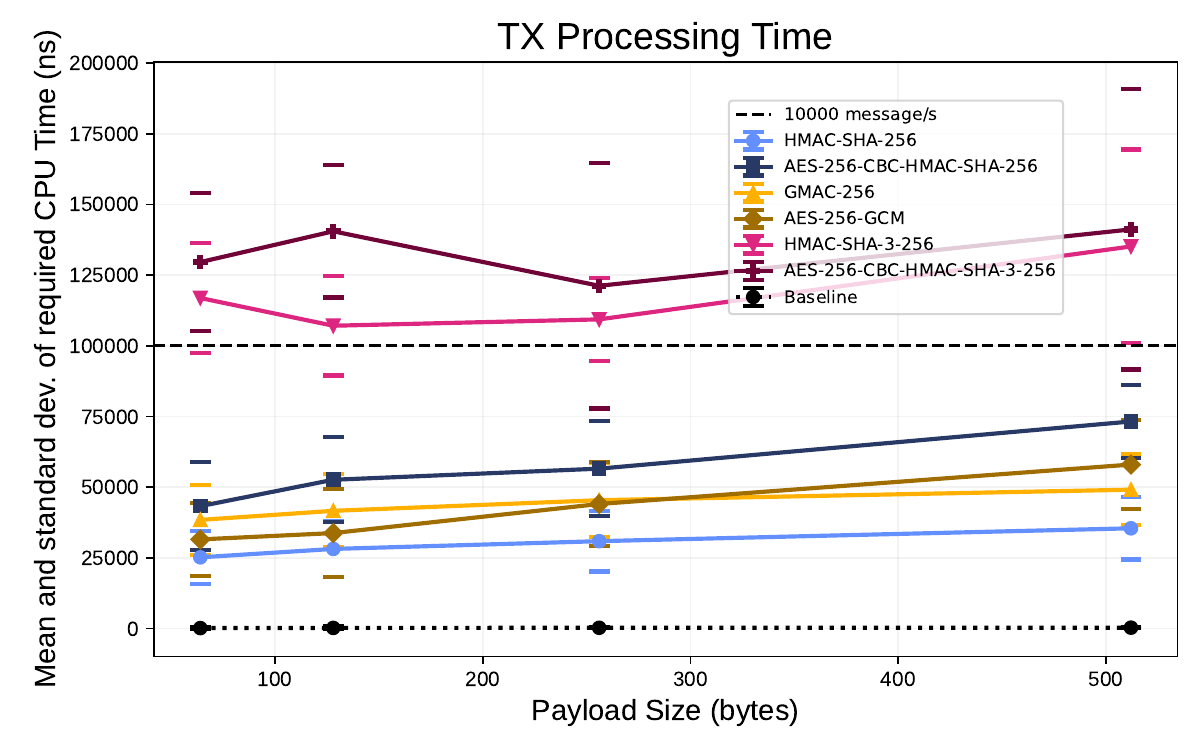}
    \caption{SecDT transmitter performance results, compared to the maximum permissible processing time for 10000 message/s and an insecure baseline.}
    \label{fig:result-bench-tx}
    
\end{figure}

\begin{figure} [ht]
    \centering
    \includegraphics[width=0.85\linewidth,height=0.95\textheight,keepaspectratio]{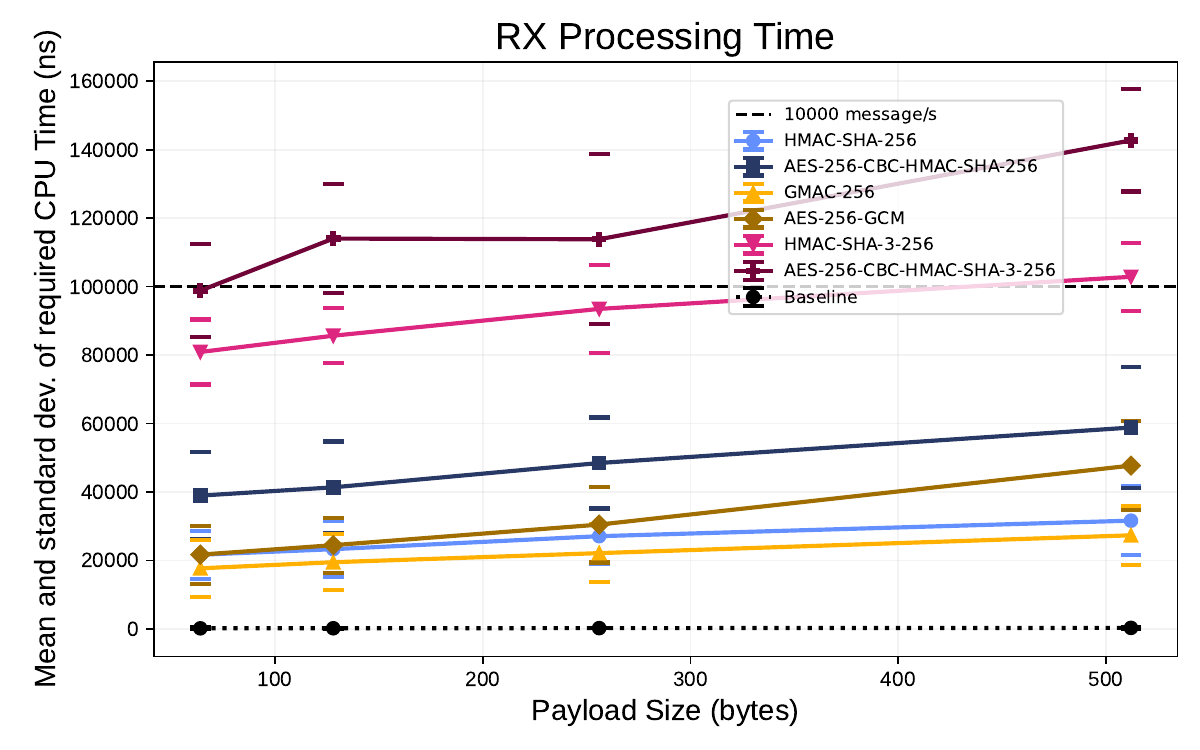}
    \caption{SecDT receiver performance results, compared to the maximum permissible processing time for 10000 message/s and an insecure baseline.}
    
    \label{fig:result-bench-rx}
\end{figure}

Figures \ref{fig:result-bench-tx} and \ref{fig:result-bench-rx} show that all evaluated security profiles exceed the target rate of 5000 messages per second with considerable margin. In order to aid clarity the figures compare the measured processing times against a stricter 10000 messages per second threshold.

<<<<<<< HEAD
As expected, receiver-side processing requires more CPU time than transmitter-side processing when confidentiality protection is enabled, as decryption and authentication verification must both be performed. The noticeably slower performance of security profiles using SHA3 based HMAC has been determined to have its source in the underlying mbedTLS library implementation.
=======
As expected, receiver-side processing requires more CPU time than transmitter-side processing when confidentiality protection is enabled, as decryption and authentication verification must both be performed. The noticeably larger performance hit on security profiles employing SHA-3 hashing has been determined to be caused by mbedTLS implementation specifics after specific benchmarking.
>>>>>>> 69-sign-iv

While SecDT's effect is clear, it is not large enough to deteriorate TRDP communication rates for a reasonably busy ComID.

\section{Conclusions and Future Work}
\label{conclusions_and_future_work}
This paper presented SecDT, a profile-based security layer designed for TRDP communications in TCNs. SecDT introduces message-level authentication, integrity protection and confidentiality through configurable security profiles, while remaining backward-compatible with existing TRDP deployments. Some security profiles have been defined, but custom profiles may be introduced to address future security requirements.

To support secure deployment, SecDT integrates with a centralized OKMS responsible for security profile assignment, key distribution and management. The use of HTTP/2 over mTLS and SSE enables secure profile negotiation and dynamic updates. Furthermore, future improvements to TLS may be adopted by the OKMS communication channel without requiring modifications to either TRDP or SecDT.

Experimental evaluation showed that the computational overhead introduced by SecDT remains compatible with real-time requirements typical in rolling stock networks. While even the most demanding security profiles remained comfortably above the target message rate, the advantages of cryptographic agility are highlighted by the lighter workloads of simpler security profiles.

Future work includes developing and testing more security profiles, based on different algorithms and employing different key sizes, tests on more types of hardware such as low-power microcontrollers or devices with dedicated hardware acceleration, and validation with traffic more akin to the real traffic used aboard rolling stock.

\section*{Acknowledgment}

This work was supported by Construcciones y Auxiliar de Ferrocarriles, S.A. (CAF) and by the Basque Government through the Project “Intelligent, Resilient, and Deterministic Beyond-5G Networks for Critical Applications in Dynamic Scenarios” (6G-BAIZTA) of the ELKARTEK Program under Grant KK-2026/00095.

\bibliographystyle{plain} 
\bibliography{refs} 
\end{document}